# Techno-Economic Assessment of Wind-Powered Green Hydrogen Production for US Industrial Decarbonization

Bibish Chaulagain [a], Sanjeev Khanna [a]

[a] Department of Mechanical and Aerospace Engineering, University of Missouri-Columbia

**Abstract**

This study evaluates the techno-economic feasibility of supplying industrial thermal loads with green hydrogen produced via water electrolysis using two pathways off-grid systems powered by co-located wind turbines and battery energy storage (BESS), and on-grid systems that procure electricity directly from the wind farm power node and operate electrolysers in response to real-time locational marginal prices (LMPs).The optimization results show that off-grid wind-to-hydrogen configurations in high-resource regions can achieve levelized costs of hydrogen (LCOH) on the order of $7/kg, driven by high wind capacity factors and optimized BESS sizing that ensures operational continuity .Similarly in, on-grid, price-responsive operation achieves LCOH values of $0.5/kg, reflecting sensitivity to electricity market volatility. Overall, the results suggest that Midwest wind-rich regions can support competitive green hydrogen production for industrial heat, with grid-connected electrolysers remaining attractive in locations with frequent low LMP periods. This dual-path analysis provides a transparent framework for industrial hydrogen deployment and highlights practical transition strategies for decarbonizing U.S. manufacturing.

**Keywords: PEM water electrolysis, Green hydrogen; decarbonization , Real-time locational marginal prices**

## 1. Introduction

Natural gas remains a cornerstone of the U.S. energy landscape, serving as a vital fuel for heating, and industrial processes. In 2024, U.S. natural gas consumption reached a record average of 90.3 billion cubic feet per day (Bcf/d), marking a 1% increase from 2023 and setting new monthly highs in both winter (January) and summer (July)[1]. The industrial sector, which accounts for approximately 33-35% of total U.S. natural gas use, consumed around 22-25 Bcf/d in 2024, holding steady year-over-year despite fluctuations in other sectors[2]. Forecasts indicate further growth, with the U.S. Energy Information Administration (EIA) projecting a new record of 91.4 Bcf/d in 2025, driven by increases across residential, commercial, and industrial sectors.[3]

However, this reliance comes at a significant environmental cost. Combustion of natural gas in the industrial sector contributed substantially to U.S. energy-related CO2 emissions, which totaled approximately 4,777-4,790 million metric tons (MMmt) [4]. Industrial direct emissions is responsible for about 23-30% of total U.S. greenhouse gas emissions when including indirect electricity-related impacts[5]. Small and medium-sized manufacturers (SMMs), in particular, rely heavily on natural gas for high-temperature thermal processes in industries analyzed in recent studies[6]. Globally, industrial emissions stand at 9 GtCO2 annually[7], underscoring the urgency for decarbonization to meet Sustainable Development Goals requiring reductions by 45% GtCO2 of 2010 by 2030.[8]

To address these emissions challenges, a broader shift toward renewable energy sources like solar, wind, and others is crucial across all sectors. While direct electrification can cover many energy needs, specific industries that rely on natural gas for high-temperature processes require a dedicated alternative; green hydrogen offers a versatile, carbon-free solution for these hard-to-abate industrial applications.[9] [10]. The global green hydrogen market, valued at USD 7.98 billion in 2024, is projected to reach USD 11.86 billion in 2025 and surge to USD 60.56 billion by 2030, growing at a CAGR of 38.5%[11]. Alternative forecasts project the total market value will grow with a compound annual growth rate (CAGR) of 67.19% from 2023 to 2033, reaching an estimated USD 141.29 billion by 2033[12].

A viable pathway for industrial decarbonization involves leveraging surplus renewable electricity from intermittent sources like wind to produce hydrogen through water electrolysis : a mature technology using either alkaline electrolyzers or proton-exchange-membrane (PEM) with efficiencies typically up to 80% [13]. This process generates green hydrogen $H_2$ as a versatile, storable, and zero-carbon fuel alternative to natural gas in high-temperature applications across heavy industries, including steel, cement, chemicals, and glass manufacturing [14]. The synergy between renewable generation, hydrogen production, and end-use creates a robust, zero-carbon energy cycle for industrial consumers, enhancing their operational flexibility and long-term energy security[15]. Hydrogen's applicability to both high and medium temperature processes means it can directly displace carbon-intensive fuels in sectors like steel manufacturing[16]. This fuel switch, while promising, necessitates technical adaptations of current equipment. Implementing hydrogen in industrial settings frequently requires specific engineering changes, such as optimizing burner designs and adjusting control systems for air/fuel management.[17] [18] [19].

Regions with high wind penetration, Texas (approximately 120 TWh annually), Oklahoma (approximately 40 TWh in 2024), Kansas (over 30 TWh in 2024), Iowa (over 35 TWh in 2024), and Illinois (approximately 18 TWh in 2024) demonstrate the reliability and significant contribution of wind power to the grid [20].These high-penetration areas frequently experience periods where high wind generation drives locational marginal prices (LMPs) to zero or negative levels, creating operational windows for processes like green hydrogen production by using electrolyzer at near-zero or negative electricity costs[21].

The Industrial Assessment Centers (ITAC) program of the U.S. Department of Energy (DOE) has been conducting assessments in Small and Medium-sized Enterprises (SMEs) to improve energy efficiency and promote the use of renewable energy sources. Building on this initiative, the present study focuses on replacing heat applications in various industries using hydrogen produced from wind energy. This approach is particularly relevant in high-penetration regions, such as those within Texas, Oklahoma, Kansas, Iowa, and Illinois. These areas feature substantial wind energy generation capacity and frequently experience periods where high wind generation drives locational marginal prices (LMPs) to zero or negative levels. This research conducts a comprehensive analysis of on-grid and off-grid wind energy systems, integrating real-time LMP pricing data and existing wind energy harvesting potential to determine the optimal sizing and operational strategy for an electrolyzer system. The optimization demonstrates a viable pathway for green hydrogen integration into industrial heat processes based on real-world market price signals.

## 2. Literature review

Various attempt of optimization has been done for generation of green hydrogen. Koutroulis et al. employed Genetic Algorithms (GA) to minimize the 20-year total system cost for stand-alone PV/wind-generator systems.[22] Similarly, Ekren and Ekren used the Simulated Annealing (SA) algorithm to find the optimal PV size, wind turbine rotor swept area, and battery capacity for hybrid energy system total cost minimization[23] .Borowy and Salameh developed a Minimum Cost-Based Sizing Methodology for calculating the optimum size of battery banks and PV arrays based on the minimum total cost for a desired system reliability [24]. Focusing on configuration and reliability, Yang et al. created the Hybrid Solar-Wind System Optimization Sizing (HSWSO) Model to optimize the capacity sizes of different components (wind turbines, PV panels, battery banks)[25]; Huneke et al. used Linear Programming (LP) methods for the optimal mix/configuration of solar, wind, storage, and a diesel generator set[26]; and Chedid and Rahman also employed Linear Programming (LP) for unit sizing and control to minimize the average production cost of electricity and monitor energy management [27] .Furthermore, Nallolla and Perumal utilized HOMER software for simulation and optimization as part of a Techno-Economic Analysis to determine the optimal design of a hybrid off-grid system[28]. The optimization efforts also extended to the growing field of green hydrogen production: Özgür and Mert used Response Surface Methodology (RSM) to achieve the maximum hydrogen production efficiency by optimizing electrolysis time, voltage, and catalyst amount [29]. Ahmad and Yadav employed a combined approach of RSM, Genetic Algorithms (GA), and Particle Swarm Optimization (PSO) to determine the optimal operating parameters for maximizing hydrogen production through wastewater electrolysis[30] and Rezk et al. focused on Modeling and Optimization techniques with the general objective of maximizing green hydrogen production from water electrocatalysis [31]. The techno-economic aspects of hydrogen were addressed by Yan et al., who performed a Techno-economic Analysis for Sorption Enhanced Steam Methane Reforming (SE-SMR) configurations[32], and Barros Souza Riedel et al. who evaluated the technical-economic performance of a renewable hydrogen production system from Solar PV and Hydro synergy [33]. Zghaibeh et al. performed optimization through HOMER-Pro software for green hydrogen production in a combined hydroelectric-photovoltaic system, achieving an optimized Levelized Cost of Hydrogen (LCOH)[34]. On a systemic level, Kookos proposed a Novel mathematical programming formulation for the simultaneous optimal design and operation of standalone green hydrogen systems [35], Brahim and Jemni proposed a Mathematical Model to optimize hybrid systems integrating various components, accounting for component degradation [36], and Alhussan et al. applied a Hybrid Dynamic Optimization Algorithm for green hydrogen production ensemble forecasting[37]. Finally, Gorji provided a review of Metaheuristic Optimization methods for addressing challenges across the green hydrogen supply chain[38]

The various studies on green hydrogen production and adoption confirm a clear pathway toward its widespread commercial viability, highlighting crucial areas for cost optimization and market integration. The collective research establishes that the Levelized Cost of Hydrogen (LCOH) is the primary metric for economic feasibility, with its reduction being highly dependent on the cost of renewable electricity and the declining capital cost of electrolyzer technology. The economic outlook is positive, with global cost projection models indicating that the LCOH is expected to fall below $5/kg by 2030 for systems utilizing solar, onshore, and offshore wind sources. This significant cost reduction is predicated on projected declining capital costs for both Alkaline and PEM electrolyzers due to ongoing R&D and scale effects[39]. This trend is underscored by a techno-economic analysis in Poland, which projected a remarkable long-term decrease in LCOH for PEM technology due to technological learning, moving toward a highly competitive range of €1.23–2.03/kg by 2050 [40].

A techno-economic evaluation in the offshore regions of Korea comparing Alkaline Electrolyzer (AEL), Proton Exchange Membrane Electrolyzer (PEMEL), and Anion Exchange Membrane Electrolyzer (ANIONEL) electrolysis technologies confirmed that the electricity cost is the most influential variable on LCOH[41]. Similarly, a multi-objective optimization of a solar-powered green hydrogen system in Boulder, Colorado, United States, aiming to minimize LCOH and Net Present Cost (NPC), found that the overall system cost is significantly affected by the capital cost of the electrolyzer and PV system[42].System efficiency is a key area of focus, with research in the UK/Europe on a 10-MW PEM electrolyser plant demonstrating the technical viability of integrating Waste Heat Recovery (WHR) using an Organic Rankine Cycle (ORC), which increases the overall system efficiency from 71.4% to 98%, providing a clear path to enhance economic benefit[43]. Furthermore, researchers affiliated with Korea and the United States quantified the impact of electrolyzer degradation (performance loss) in an Alkaline Water Electrolysis (AEL) system, yielding essential data for developing robust maintenance strategies that protect long-term system economics[44]. The analysis in Germany comparing AEL, PEMEL, and Solid Oxide Electrolyzer (SOEL) technologies also established specific, achievable low LCOH targets for green hydrogen by 2030 and 2050 under low electricity price scenarios to successfully compete with blue and grey hydrogen[45].

The research validates green hydrogen's economic feasibility in specific regional and industrial contexts, often driven by high-quality renewable resources and targeted policy interventions, while addressing logistics. An economic model for the United Arab Emirates (UAE) projected highly competitive green hydrogen production costs between \$0.79/kg and \$1.94/kg in 2030, suggesting that this cost structure will enable cost viability in the ammonia sector by 2029 (at prices below \$1.35/kg)and necessitate a massive build-out of approximately 55 GW of electrolyzer capacity by 2050 to meet domestic industrial demand[46]. Addressing logistics, a global cost projection analysis focusing on Germany concluded that importing green ammonia is generally more cost-efficient for domestic ammonia production than importing liquid green hydrogen $LH_2$, which supports a strategic focus on hydrogen derivatives and the development of regional $H_2$ pipeline grids for cost-effective trade [47]. Resource optimization is further confirmed by a study on green hydrogen from offshore wind in the Taiwan Strait, which showed that optimizing the electrolyzer configuration (centralized vs. distributed) based on offshore distance is a key strategy for LCOH minimization[48].Additionally, a cost analysis of PEM plants in Brazil confirmed that while non-potable water sources (seawater, industrial wastewater) require careful treatment optimization, their costs are not the dominant LCOH factor, making these sources a viable option in water-stressed regions[49].Policy analysis confirmed that the European Union (EU) has a comprehensive public support strategy across the hydrogen value chain[50], while research focusing on the U.S. Inflation Reduction Act (IRA) tax credits, using California as a case study, recommends incorporating Willingness to Pay (WTP) from the consumer side into policy designs to successfully activate demand in hard-to-decarbonize sectors[51] .Finally, techno-economic assessments have validated the feasibility of green hydrogen in specific, high-impact applications, offering clear pathways for decarbonization: a study in Naples, Cassino, and Salerno, Italy, confirmed the benefits of a multi-energy system (MES) for the co-production of green hydrogen, renewable electricity, and heat, achieving competitive levelized costs of total output energy (0.175–0.183 €/kWh) [52]; a study in Türkiye confirmed the economic justification for using Hydrogen-Ammonia fuel blends in tugboat engines [53] and an assessment in the UAE validated the feasibility of a green-hydrogen-based microgrid design for commercial buildings using a solar PV system with a reversible fuel cell [54].

## 3. Methodology

This study develops a techno-economic optimization framework to assess green hydrogen as a replacement for natural gas process heat in five representative U.S. manufacturing subsectors food processing, chemical industries , glass and ceramics industry, electronics manufacturing industry, and metal industry. For each sector, the annual thermal demand is first converted to an equivalent annual hydrogen requirement using the lower heating value of $H_2$. Two complementary supply configurations are then evaluated. In the off-grid case, hydrogen is produced by a dedicated wind–battery–electrolyzer system whose design (installed wind capacity, battery power and energy ratings, and electrolyzer size) and hourly operation are co-optimized using deterministic mixed-integer linear programming (MILP) and site-specific wind-speed time series

In the grid-connected case, hydrogen is produced by an electrolyzer directly supplied from the wholesale electricity market, with real-time or day-ahead locational marginal prices (LMPs) from wind farm only nodes used to construct multiple stochastic price scenarios; a stochastic MILP is then solved in which electrolyzer capacity is chosen in the first stage and hourly dispatch, on/off status decisions are optimized in the second stage across all LMP scenarios.

In both configurations, the objective is to minimize the net present cost over a 20-year horizon including capital costs for wind turbines, batteries, and electrolyzers, annual operating and maintenance expenditures, stack replacement, burner conversion costs on the demand side, and (for grid-connected cases) and expected electricity costs subject to operational constraints on minimum load, ramp rates, and minimum up/down times and a hard constraint that total annual hydrogen production meets or exceeds the sector-specific annual demand. Hydrogen storage tanks and compression are deliberately excluded from the system boundary so that the comparison focuses on the marginal cost and sizing of generation and power-conditioning assets needed to decarbonize industrial heat, while more detailed modeling of storage technologies and siting/regulatory constraints is reserved for future work.

This study quantifies the economic and technical feasibility of replacing 25%–100% of natural gas thermal loads in small- and medium-sized manufacturers (SMMs) with grid-connected green hydrogen produced via electrolysis, leveraging real-time locational marginal prices (LMPs) from existing wind-dominated U.S. electricity markets and offgrid system. The analysis targets five high-temperature industrial sectors food processing, chemical industries , glass and ceramics industry, electronics manufacturing industry, and metal industry.and electronics & instrument industry where natural gas is used for process heating at 100–1600 °C. By exploiting negative and zero LMP hours driven by wind overgeneration in MISO Iowa the framework identifies optimal electrolyzer sizing.

To capture future price volatility, a mixed-integer linear programming (MILP) model is developed. The first stage determines the electrolyzer rated capacity (P_rated) prior to uncertainty realization. The second stage includes recourse decisions—hourly power dispatch (P_s,t), on/off status (u_s,t), hydrogen production, and inventory for each scenarios generated via multiplicative Gaussian noise applied to real current LMP data. The objective minimizes the expected net present cost (NPC) over a 20-year lifetime, incorporating CAPEX, OPEX, stack replacement, burner retrofitting, and expected electricity.

Figure 1 illustrates the research workflow. The process begins with data acquisition of SMM thermal profiles and real-time LMPs, followed by scenario generation to model price uncertainty. System component modeling defines electrolyzer physics. The core optimization solve the stochastic MILP, and post-processing computes levelized cost of hydrogen (LCOH), $CO_2$ abatement, and carbon price breakeven relative to natural gas. Each step is detailed below.

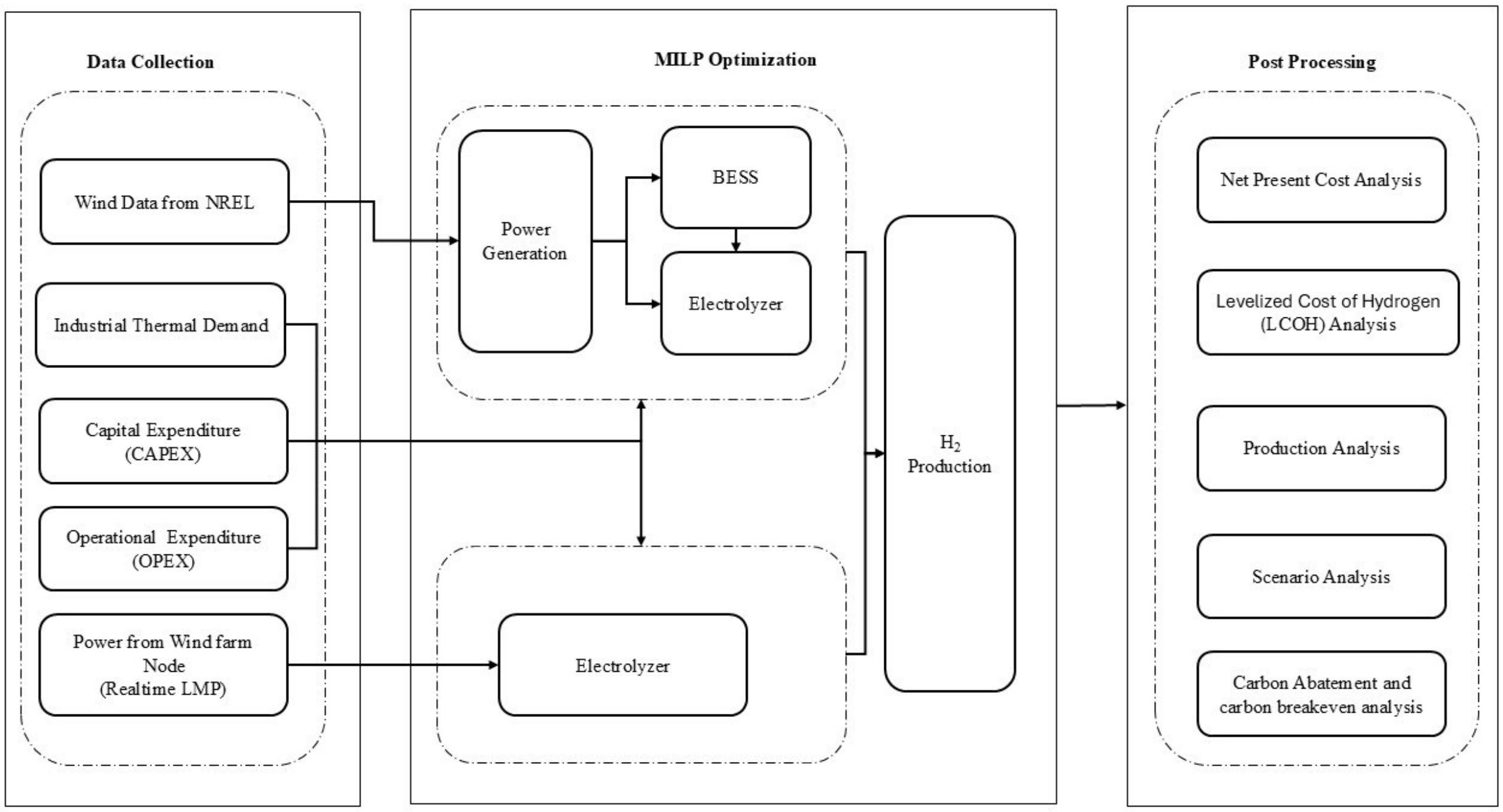

**Fig 1: Schematic overview of the methodology**

## 4. Data Acquisition

Industrial thermal demand profiles for the five-target small- and medium-sized manufacturer (SMM) sectors food processing, chemical industries , glass and ceramics industry, electronics manufacturing industry, and metal industry were compiled using data from U.S. Department of Energy Industrial Assessment Center (IAC) audits and the median annual natural gas consumption was used to represent each sector's typical demand; a choice made over the mean value to minimize the influence of potential data outliers. This median consumption was then translated into an equivalent annual hydrogen mass requirement utilizing the lower heating value (LHV) of hydrogen, standardized at 33.3 kWh/kg. (see Table 1).

Real-time locational marginal price (LMP) data was collected at hourly intervals for the entirety of calendar year 2024, sourced directly from the public market portals of MISO. Nodes were specifically selected from wind generation farms that exclusively produce wind-powered energy, in order to capture significant price volatility driven by wind generation. The selected node of the MISO, Iowa, was chosen because they frequently exhibit low LMP events indicative of wind overgeneration suitable for opportunistic, low-cost electrolysis and our Midwest Industrial Training and Assessment Center also do industrial audit in those areas.

System cost parameters were derived from recent peer-reviewed techno-economic studies and industry reports published, with future projections adjusted for inflation and learning curves. The capital expenditure (CAPEX) , Annual operational expenditure (OPEX) for the electrolyzer system derived from peer-reviewed techno-economic studies (including balance-of-plant components such as demineralized water systems, gas separators, rectifiers, and control units.

Hydrogen is produced on-site via a grid-connected electrolysis system, which removes the need for costly transportation, or liquefaction infrastructure. This co-location strategy minimizes supply chain risks and facilitates demand-responsive production, allowing electrolyzers to ramp operations in real-time to capitalize on negative LMP windows, thereby directly linking surplus renewable energy to industrial decarbonization efforts.

The cost associated with converting existing industrial burners to operate with hydrogen varies significantly across different SMM categories due to differences in process temperatures and current burner configurations. These retrofit costs are primarily dependent on the power rating of the burner being modified. For instance, industries with high-temperature requirements typically incur higher conversion costs due to the need for specialized materials and design adaptations to accommodate hydrogen's unique combustion properties. In contrast which generally operate at lower temperatures, face comparatively lower retrofit expenses. For the cost of burner replacement, we have utilized data from the ITAC database, which provides information on the implementation costs of burner replacement. We have assumed that the conversion of existing burners to those capable of operating with hydrogen will incur similar costs to those associated with complete burner replacement.

The wind speed of the location is provided by the NREL wind toolkit [55]. Though the simulated data of wind speed was available upto 2014 we assumed that same wind speed for our current period of time. And the realtime LMP was extracted from one of the MISO nodes which has electricity from Wind only.

**Table 1: Annual NG consumption in different Industrial sector.**

| **Industry** | **MMBTU** | **Operation hour** |
|---|---:|---:|
| Food Processing | 80,992 | 8760 |
| Glass & Ceramics Industry | 75,402 | 8760 |
| Electronics & Instruments | 76,100 | 8760 |
| Chemical Manufacturing | 37,651 | 8760 |
| Metals & Machinery | 30,536 | 8760 |

**Table 2: Parameters for Wind Turbine System.**

| Parameters | Value | Reference |
|---|---|---|
| Hub height (m) | 110 m | [56] |
| Wind Speed | NREL toolkit | [55] |
| Cut-in wind speed (m/s) | 4 (m/s) | [56] |
| Cut-out wind speed | 25 (m/s) | [56] |
| Rated Power( $P_r$) | 3.4 MW | [57] [56] |

**Table 3:Parameters for Battery System.**

| Parameter | Value | Reference |
|---|---|---|
| Self-discharging factor ,$\sigma$ | 0.0083% | [58] |
| Charging efficiency,$\eta_{batt,ch}$ | 98% | [59] |
| Discharging efficiency,$\eta_{batt,dch}$ | 98% | [59] |
| Minimum SOC, $SOC_{batt,min}$ | 15% | [58] |
| Maximum SOC,$SOC_{batt,max}$ | 95% | [58] |
| Initial SOC | 50 | |

**Table4: Economic Parameter**

| Parameter Value Reference | Value | Reference |
|---|---|---|
| Wind Turbine CAPEX | 1544.24$/kW | [60] |
| Wind Turbine OPEX | 45$/kW/yr | [60] |
| Battery system CAPEX | 276$/kW | [61] |
| Battery system OPEX | 2.5%ofCAPEX | [62] |
| Battery replacement cost | 10%ofCAPEX | [61] |
| Electrolyser system CAPEX | 705$/kW | [63] |
| Electrolyser system OPEX | 7.5%ofCAPEX | [64] |
| Stack replacement cost | 35%ofCAPEX | [65] |
| System installation cost | 12%ofoverallCAPEX | [65] |
| Discount factor | 10% | [65] |
| System life time | 20years | |
| | | |

## 5. System Modeling and Optimization

This study formulates mixed-integer linear programming (MILP) models to minimize the net present cost (NPC) of green hydrogen production for industrial process heat, under both off-grid (wind + battery + electrolyzer) and on-grid (grid-connected electrolyzer) configurations. In both cases, hydrogen is produced solely via water electrolysis using PEM and used to replace natural gas (NG) in five energy-intensive manufacturing. Sector-specific annual thermal demands are converted to equivalent hydrogen mass requirements, and the optimization co-designs the energy supply, electrolyzer size, and hourly operating schedule subject to realistic technical and economic constraints.

Hydrogen storage tanks are explicitly excluded from the optimization. This reflects both the added physical and regulatory complexity of gaseous hydrogen storage (pressure vessels, safety distances, permitting) and the focus of this work on the generation side (wind/BESS/grid + electrolyzer) rather than storage system design. Instead, the models enforce that annual hydrogen production meets or exceeds the annual thermal hydrogen requirement, and operational reliability is ensured through appropriate sizing of the wind farm and battery (off-grid case) or by relying on firm grid supply (on-grid case).

For each industrial sector $j$, the annual thermal demand $E_j^{\mathrm{th}}$(MWh $_{\mathrm{th}}$/year) is taken from ITAC data shown in Table 1. The **annual hydrogen demand** (kg/year) is then

$$D_j^{\mathrm{H_2}} = \phi \frac{E_j^{\mathrm{th}}}{\mathrm{LHV}_{\mathrm{H_2}}^{\mathrm{kWh}}} \qquad (1)$$

where $\mathrm{LHV}_{\mathrm{H_2}}^{\mathrm{kWh}} \approx 33.3$ kWh/kgis the lower heating value of hydrogen.

In both configurations, hydrogen is produced by a single aggregated electrolyzer with rated capacity $P_{\max}$(MW). For the off-grid case we treat $P_{\max}$as a design variable; for the on-grid case we call it $P^{\mathrm{rated}}$. The instantaneous electrolyzer power in hour $t$(and scenario $s$ in the stochastic model) is $P_t$or $P_{s,t}$(MW). Assuming constant electrical efficiency $\eta_{\mathrm{el}}$and using the higher heating value (HHV) of hydrogen $\mathrm{HHV}_{\mathrm{H_2}}$(J/kg), the mass of hydrogen production per hour is

$$H_t^{\mathrm{prod}} = \frac{\eta_{\mathrm{el}} P_t}{\mathrm{HHV}_{\mathrm{H_2}}} \qquad (2)$$

Electrolyzer operation is controlled by **binary on/off variables** and a **minimum load constraint**:$x_t \in \{0,1\}$ with ramping factor of adapted which is generally adopted in PEM electrolyzer dynamics.

The rated electrolyzer capacity $P_{\max}$is a decision variable, so the on/off product $x_t P_{\max}$would be bilinear. To keep the problem linear, we introduce an auxiliary variable $P_t^z$(MW) that represents the **available electrolyzer headroom when ON**, and apply a standard big-M / McCormick linearization:

$$\begin{aligned} P_t^z &\le P_{\mathrm{el.max}} - (1 - x_t)\, P_{\mathrm{el.min}} \\ P_t^z &\ge P_{\mathrm{el.max}} - (1 - x_t)\, P_{\mathrm{el.max}} \\ P_{\mathrm{el.min}}\, x_t \le\ & P_t^z \le P_{\mathrm{el.max}}\, x_t \end{aligned} \qquad (3)$$

where $P_{\min}$and $P_{\max}$are lower and upper design bound for electrolyzer size. The actual operating power is then constrained between a minimum part-load and full available capacity:

$$y_{\min} P_t^z \le P_t \le y_{\max} P_t^z, \qquad (4)$$

with $0 < y_{\min} \le y_{\max} \le 1$. This structure closely follows MILP formulations used in the literature for flexible electrolyzer scheduling[66].

### 5.1 Off-grid model

The off-grid configuration is powered by a dedicated wind farm and a battery energy storage system (BESS). Decision variables includes $P_{\mathrm{rated}}^{\mathrm{WT}}$(MW); installed wind capacity,$E^{\mathrm{batt}}$(kWh): BESS energy capacity and $P_{\mathrm{rated}}^{\mathrm{batt}}$(MW): BESS power capacity.

For each hour, the available wind power per unit capacity is given by a three-region power curve $p^{\mathrm{WT}}(v_t)$(MW/MW):[67]

$$p^{\mathrm{WT}}(v_t) = \begin{cases} 0, & v_t < v_{\mathrm{ci}} \text{ or } v_t \ge v_{\mathrm{co}}, \\ \dfrac{P_{\mathrm{r}}}{v_{\mathrm{r}}^3 - v_{\mathrm{ci}}^3}\,(v_t^3 - v_{\mathrm{ci}}^3), & v_{\mathrm{ci}} \le v_t < v_{\mathrm{r}}, \\ P_{\mathrm{r}}, & v_{\mathrm{r}} \le v_t < v_{\mathrm{co}}, \end{cases} \qquad (5)$$

where $v_{\mathrm{ci}}, v_{\mathrm{r}}, v_{\mathrm{co}}$are the cut-in, rated, and cut-out speeds, and $P_{\mathrm{r}}$ is the reference power.

$$P_t^{\mathrm{WT}} = A_{wt}\, p^{\mathrm{WT}}(v_t) \eta_{wt} \qquad (6)$$

The BESS is modeled with separate charge and discharge power variables $P_t^{\mathrm{ch}}$and $P_t^{\mathrm{dch}}$(MW) and an energy state of charge $E_t^{\mathrm{batt}}$(kWh). The energy balance while charging is

$$E_{t+1}^{\mathrm{batt}} = (1 - \sigma)\, E_t^{\mathrm{batt}} + \eta^{\mathrm{ch}} P_t^{\mathrm{ch}} \qquad (7)$$

The energy balance while discharging is

$$E_{t+1}^{\text{batt}} = (1-\sigma)\, E_t^{\text{batt}} - \frac{1}{\eta^{\text{dch}}} P_t^{\text{dch}} \qquad (8)$$

The total energy balance equation is given by combining both of above as

$$E_{t+1}^{\text{batt}} = (1-\sigma)\, E_t^{\text{batt}} + \eta^{\text{ch}} P_t^{\text{ch}} - \frac{1}{\eta^{\text{dch}}} P_t^{\text{dch}} \qquad (9)$$

For limiting the battery energy state within the Minimum and maximum threshold, we implement the constraint

$$SOC_{\text{batt,min}} \cdot E_{\text{batt,rated}} \leq E_t^{\text{batt}} \leq SOC_{\text{batt,max}} \cdot E_{\text{batt,rated}}$$

Where $SOC_{\text{batt,min}}$ is the allowed minimum state of charge and $SOC_{\text{batt,max}}$ is the allowed maximum state of charge of the battery.

### 5.2 Grid-connected model

In the **on-grid** configuration, the electrolyzer is supplied directly from the transmission grid and sees a time-varying electricity price given by the node-specific LMP. To reflect price uncertainty, we construct $S$equiprobable scenarios from the historical LMP time series. For each scenario $s$,

$$\pi_{s,t} = \pi_t^{\text{base}}\,(1+\epsilon_{s,t}), \epsilon_{s,t} \sim \mathcal{N}(0, \sigma_\pi^2), \qquad (11)$$

with $\sigma_\pi \approx 0.15$(15% volatility), and then convert to $/kWh for use in the cost function. This multiplicative noise preserves the sign and relative pattern of LMPs and is consistent with stochastic LMP modeling in hydrogen planning and fuel-station studies. [66]

The grid-connected problem is formulated as a two-stage **stochastic MILP**:**First stage** choose electrolyzer size, common to all scenarios, and in s**econd stage**, for each scenario $s$, choose hourly $P_{s,t}$ and $H_{s,t}^{\text{prod}}$, satisfying the operational constraints.

### 5.3 Cost model and optimization objective

The cost model is identical in structure for both configurations, with additional CAPEX/OPEX terms for wind and BESS in the off-grid case.

As Off-grid wind generation is assumed to have zero marginal cost; there is no grid electricity purchase but on On-grid the expected annual electricity cost is

$$C_{\text{year}}^{\text{elec}} = \sum_{s=1}^{S} \quad p_s \sum_{t=1}^{T} \quad \pi_{s,t}\, P_{s,t} \cdot 10^3 \Delta t, \qquad (12)$$

where $p_s = 1/S$for equiprobable scenarios and $\pi_{s,t}$is in \$/kWh.

For a project lifetime $N$years and discount rate $r$, the **present value factor** for an annual recurring cost is

$$\text{PV}(r, N) = \sum_{y=1}^{N} \quad \frac{1}{(1+r)^y} \qquad (13)$$

The total NPC combines upfront CAPEX, present value of OPEX and electricity purchases, stack/battery replacements

$$Total_{\text{OPEX}}^{\text{offgrid}} = \text{PV}(r, N)\,(C_{\text{OPEX,year}}^{\text{el}} + C_{\text{OPEX,year}}^{\text{WT}} + C_{\text{OPEX,year}}^{\text{batt}})$$

$$\text{NPC}^{\text{off grid}} = C_{\text{CAPEX}}^{\text{el}} + C_{\text{CAPEX}}^{\text{WT}} + C_{\text{CAPEX}}^{\text{batt}} + C_{\text{CAPEX}}^{\text{burner}} + Total_{\text{OPEX}}^{\text{offgrid}} + C_{\text{PV}}^{\text{stack}} + C_{\text{PV}}^{\text{batt, repl}}. \qquad (14)$$

$$\text{NPC}^{\text{on grid}} = C_{\text{CAPEX}}^{\text{el}} + C_{\text{CAPEX}}^{\text{burner}} + \text{PV}(r, N)\,(C_{\text{OPEX,year}}^{\text{el}}) + C_{\text{PV}}^{\text{stack}}. \qquad (15)$$

Assuming the optimized annual hydrogen production is approximately constant across years,the **levelized cost of hydrogen** is computed as

$$\text{LCOH} = \frac{\text{NPC}}{Q_{\text{PV}}^{\text{H}_2}} \, [\$/\text{kg}], \qquad (16)$$

evaluated separately for the off-grid and on-grid configurations. The MILP (deterministic off-grid or two-stage stochastic on-grid) minimizes NPC subject to all physical and operational constraints described above, while ensuring that annual hydrogen production meets or exceeds the NG-equivalent thermal demand for each industrial sector.

## 6.Results

### 6.1 Net Present Cost Analysis

Figure 2 presents the net present cost (NPC) of the complete green hydrogen system as a function of the natural-gas-to-hydrogen replacement ratio (25 %, 50 %, 75 % and 100 %) for the five SMM sectors in Iowa under both on-grid and off-grid configurations. The results reveal two strikingly different economic regimes that powerfully shows the transformative potential of green hydrogen when produced in wind-rich electricity markets.

For the on-grid case, which exploits real-time locational marginal prices at the Iowa node the NPC rises only modestly and sub-linearly with the replacement ratio, from approximately \$0.44–0.83 million at 25 % replacement to \$1.2–3.3 million at full 100 % replacement, depending on the sector. This gentle increase reflects strong economies of scale: the electrolyser is already heavily utilised during low price periods even at low replacement ratios, so expanding hydrogen demand primarily leverages existing capacity and cheap (often paid-to-take) electricity rather than requiring proportionally larger capital investment. Consequently, the levelised cost of hydrogen (LCOH) falls continuously with higher replacement ratios, achieving its minimum at 100 % substitution in five sectors. Full decarbonisation of the thermal load therefore emerges not as an expensive "luxury" option, but as the economically optimal outcome.

In contrast, the off-grid configuration, which relies on dedicated wind turbines sized to meet the same hydrogen demand without grid access, exhibits an almost perfectly linear NPC increase with replacement ratio, ranging from 4.4–12.1 million USD at 25 % to 19.3–42.4 million USD at 100 %. Here the capital cost of the wind farm dominates, and every additional kilogram of hydrogen demands a proportionate increase in installed wind capacity.

These results deliver a clear and policy-relevant message: in regions with mature, wind-rich wholesale electricity markets such as Iowa, grid-connected electrolysis transforms green hydrogen from a niche decarbonisation option into the lowest-cost pathway for eliminating process-related $CO_2$ emissions from high-temperature industrial heat. Achieving 100 % replacement is not only technically feasible but economically superior, delivering annual $CO_2$ savings of 3,573–9,476 tons per site (35,180 tons across all five sectors) at the lowest possible levelised cost and without requiring new dedicated renewable generation. Off-grid solutions, while viable in remote locations, impose a severe economic penalty that renders large-scale hydrogen adoption in SMMs impractical in the presence of a decarbonised grid. Green hydrogen produced via intelligent grid interaction in wind-rich nodes therefore represents one of the most immediately deployable and cost-effective deep-decarbonisation levers available to the hard-to-abate industrial sector today.

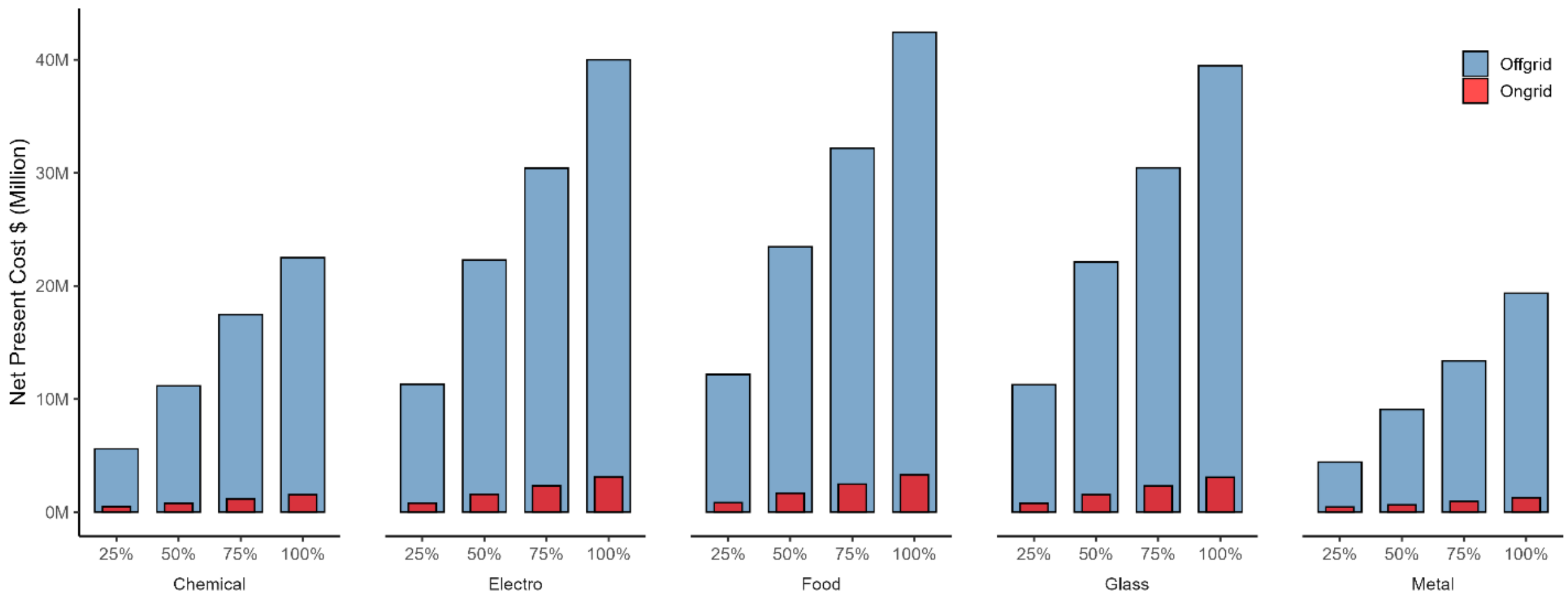

**Fig:2 NPC scenario for both on grid and off grid system in Different Industries**

## 6.2 Levelized Cost of Hydrogen (LCOH ) Analysis .

As we can see that food processing industry has the highest thermal demand and NPC cost among other category we have chosen that category for different analysis. The figure 3 presents the Levelized Cost of Hydrogen (LCOH, $/kg) for both off-grid and on-grid electrolyzer systems at different hydrogen replacement fractions in a food processing industry .For the off-grid system, which integrates wind turbines, battery storage, and an electrolyzer, the LCOH varies from 6.99 to 8.03 $/kg depending on the fraction of hydrogen demand replaced. Notably, the lowest LCOH of 6.99 $/kg occurs at full replacement (100%), while partial replacement (25–50%) results in slightly higher costs due to underutilization of the renewable generation.

In contrast, the on-grid system, which uses grid electricity with time-varying electricity prices and an electrolyzer, exhibits much lower LCOH values, ranging from 0.54 to 0.55 $/kg. Here, the LCOH decreases slightly as the hydrogen replacement fraction increases, indicating more efficient use of electricity and system operation at higher production levels.

Overall, the comparison clearly shows that the off-grid system has significantly higher hydrogen production costs due to the need for capital-intensive infrastructure (wind turbines and battery storage) to ensure continuous hydrogen supply, whereas the on-grid system leverages lower-cost grid electricity, resulting in more economical hydrogen production across all replacement fractions.

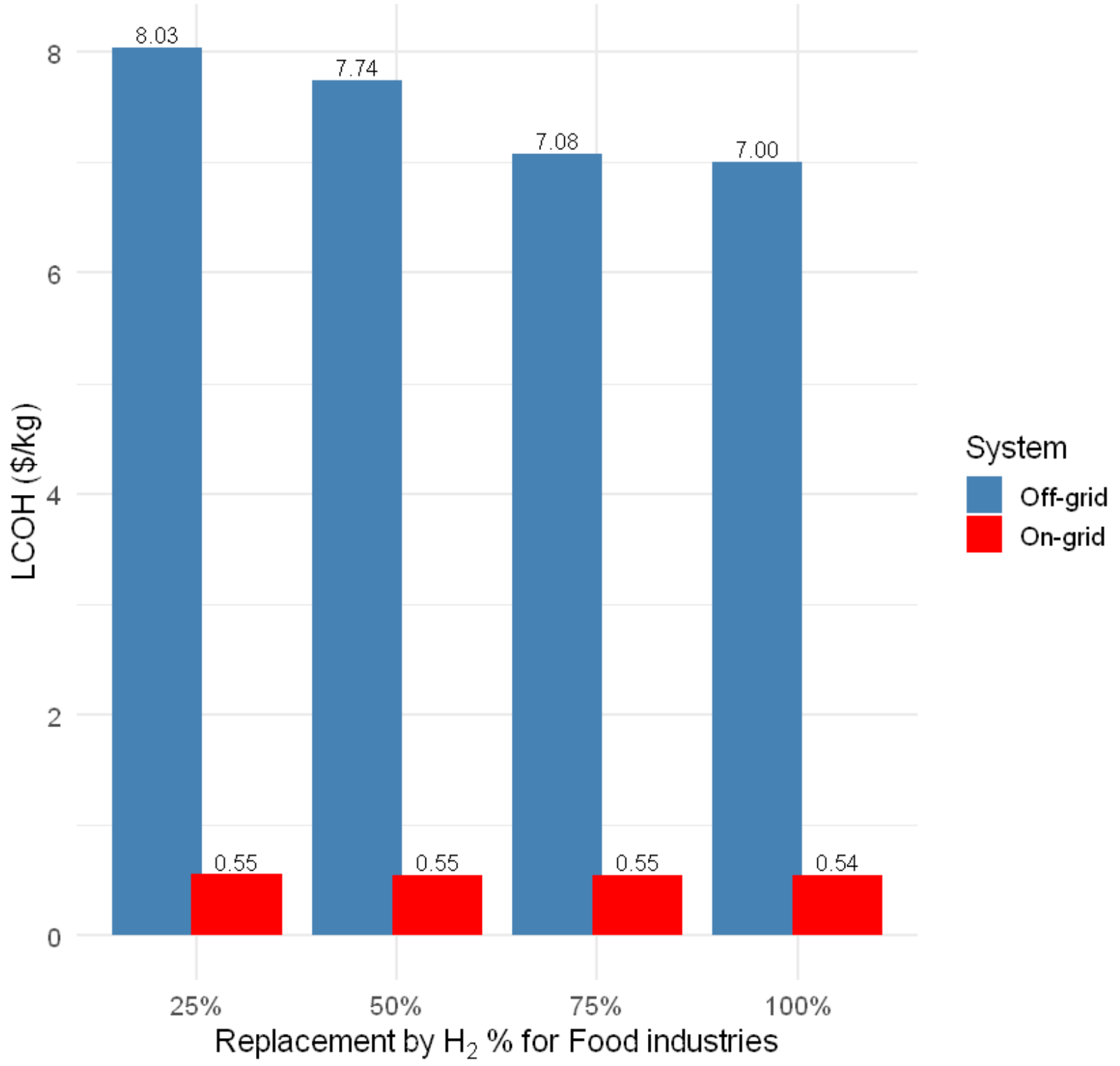

**Fig:3 LCOH for both on grid and off grid system in food processing industries.**

## 6.3 $H_2$ Production Analysis

The figure 4 below clearly shows distinct production characteristics between the off-grid and on-grid hydrogen systems. The off-grid system exhibits much higher variability and generally higher peaks in monthly hydrogen production because its output is directly dependent on wind availability, which naturally fluctuates across the year. This is visible through the wide values (e.g., 142 kg in January, 171 kg in March) and high maximum values consistently reaching 195 kg in some days.

In contrast, the on-grid system demonstrates extremely stable and controlled hydrogen production across all months. Median and upper-quartile values are nearly constant at 101 kg, with narrow variation, indicating minimal variability. This steadiness arises because the on-grid electrolyzer primarily responds to electricity price signals (LMP) rather than weather conditions. Minimum values tend to be around 10.1 kg, showing the electrolyzer rarely shuts down and operates more predictably.

Overall, the figure highlights that the off-grid system provides higher potential output but with large fluctuations, while the on-grid system offers lower but consistent hydrogen production throughout the year. These results highlight the impact of grid support on reducing operational variability and ensuring a more stable hydrogen supply, while off-grid systems must manage larger fluctuations due to reliance on intermittent renewable sources.

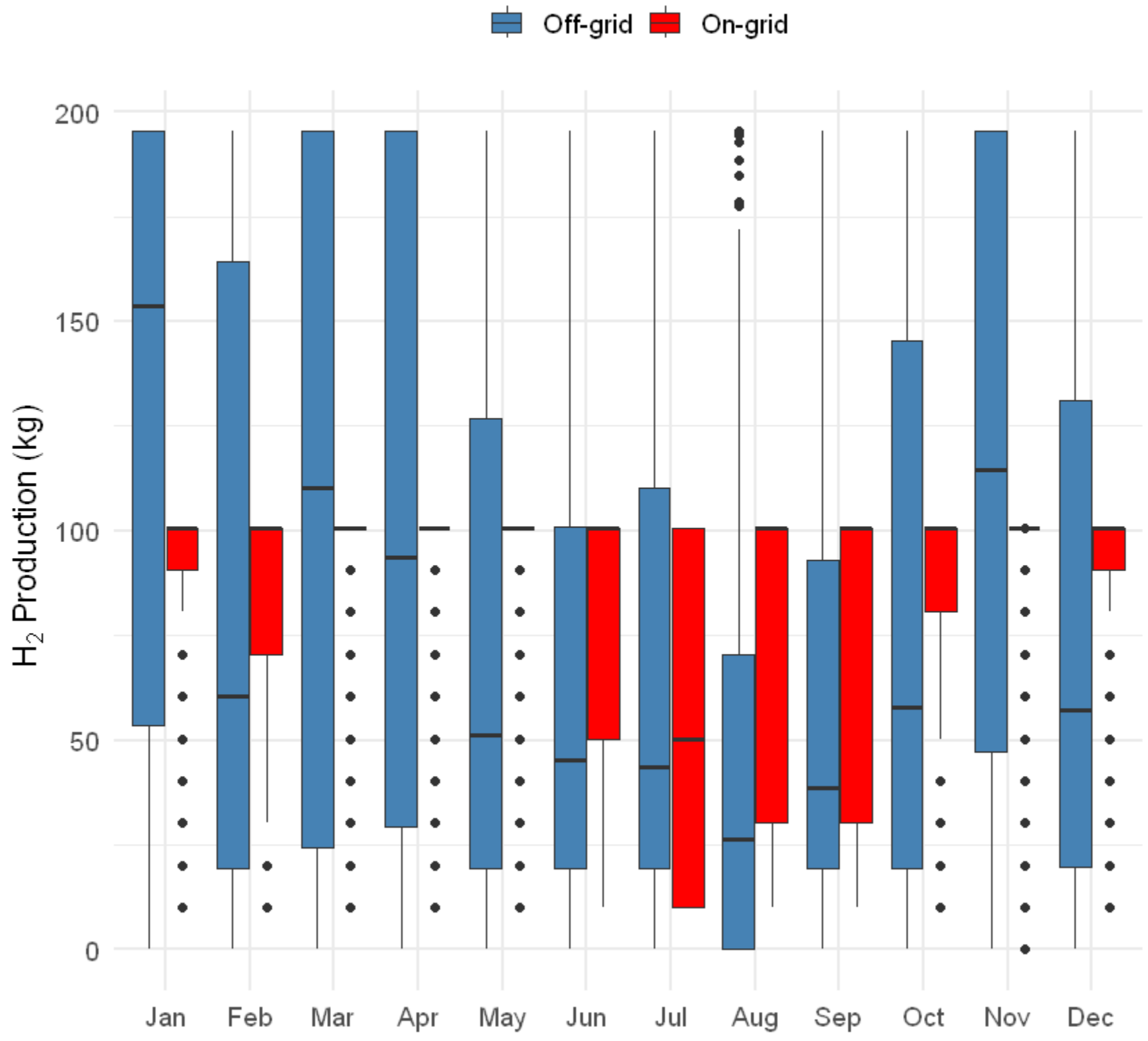

**Fig:4 Production of $H_2$ in both on grid and off grid system in food processing industries**

The distribution of the wind power is showing in the figure 5: which shows the hourly variation of the total power generated from WT in off-grid systems.The hourly wind speed at the hub height at the location of the Fort Madison is obtained form the NREL tool. We can also see the yearly H2 production pattern throughout the year when the NG is replaced by H2 by the Wind turbine +BESS system. The H2 production pattern exhibits the similar trend with the wind pattern throughout the year .

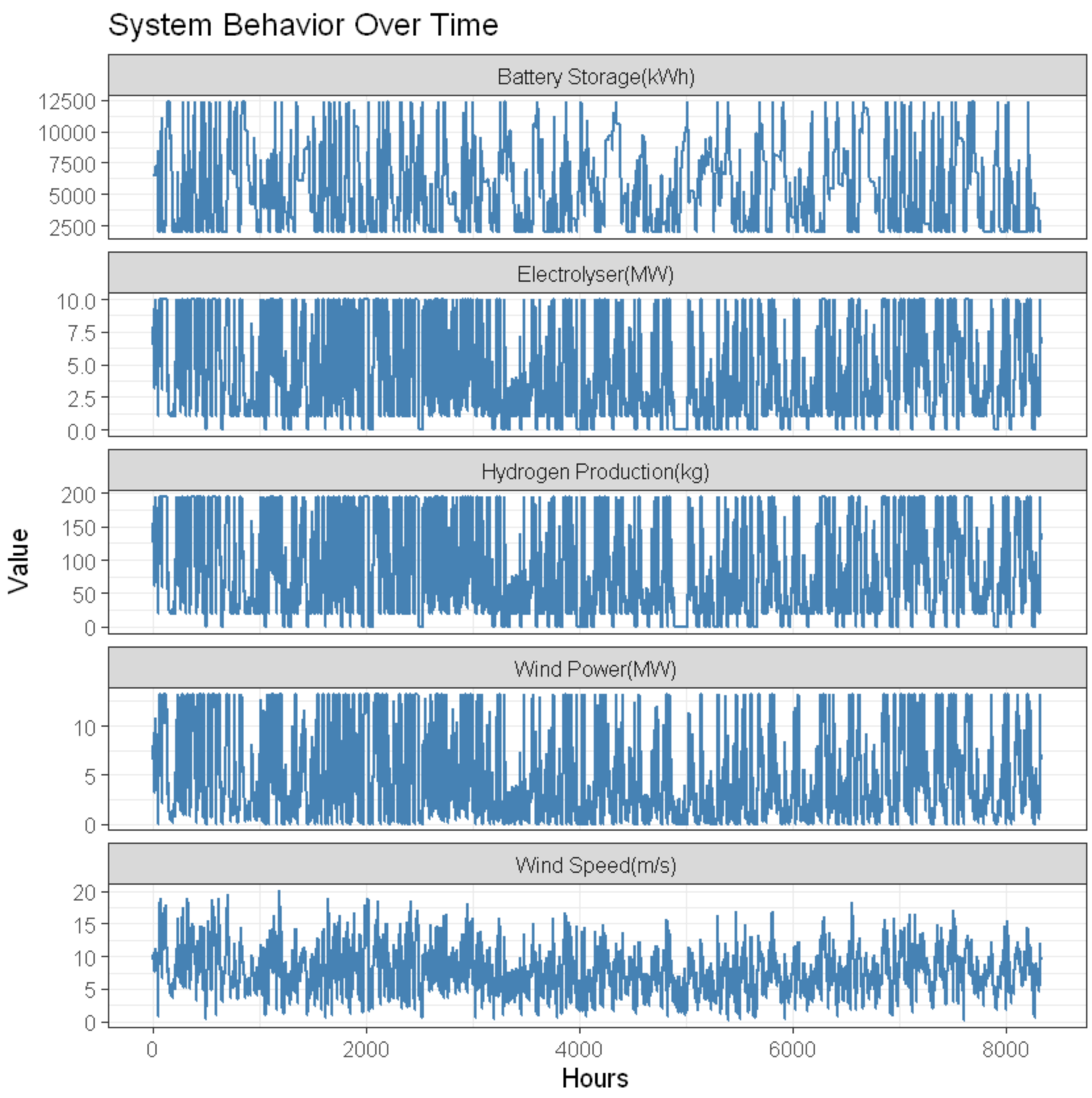

**Fig5: Electrolyzer performance, H2 production , Electrolyzer power, Wind power, Speed**

**6.4 Scenario Analysis:**

The LCOH-based analysis reveals fundamentally different cost drivers for the off-grid and on-grid hydrogen production systems. In the off-grid configuration, the LCOH is dominated by Wind Turbine CAPEX, which accounts for more than half of the total cost (49.6%), highlighting the strong dependence of off-grid hydrogen economics on large-scale Wind Turbine. Operational Expenditure represents the second major contributor at 24.1%, followed electrolyzer CAPEX while stack replacement and burner retrofit costs have only marginal influence. This indicates that cost reductions in renewable infrastructure would yield the greatest improvements in off-grid economic performance. In contrast, the on-grid system is overwhelmingly driven by electricity expenditure, which constitutes 44% of total LCOH, making the system primarily electricity-price-limited. Electrolyzer CAPEX (30.7%) and OPEX (19.6%) play secondary roles, whereas burner and stack replacement costs contribute minimally. Overall, the comparison shows that off-grid hydrogen cost is most sensitive to renewable energy and infrastructure investments, whereas on-grid hydrogen cost is dominated by electricity price.

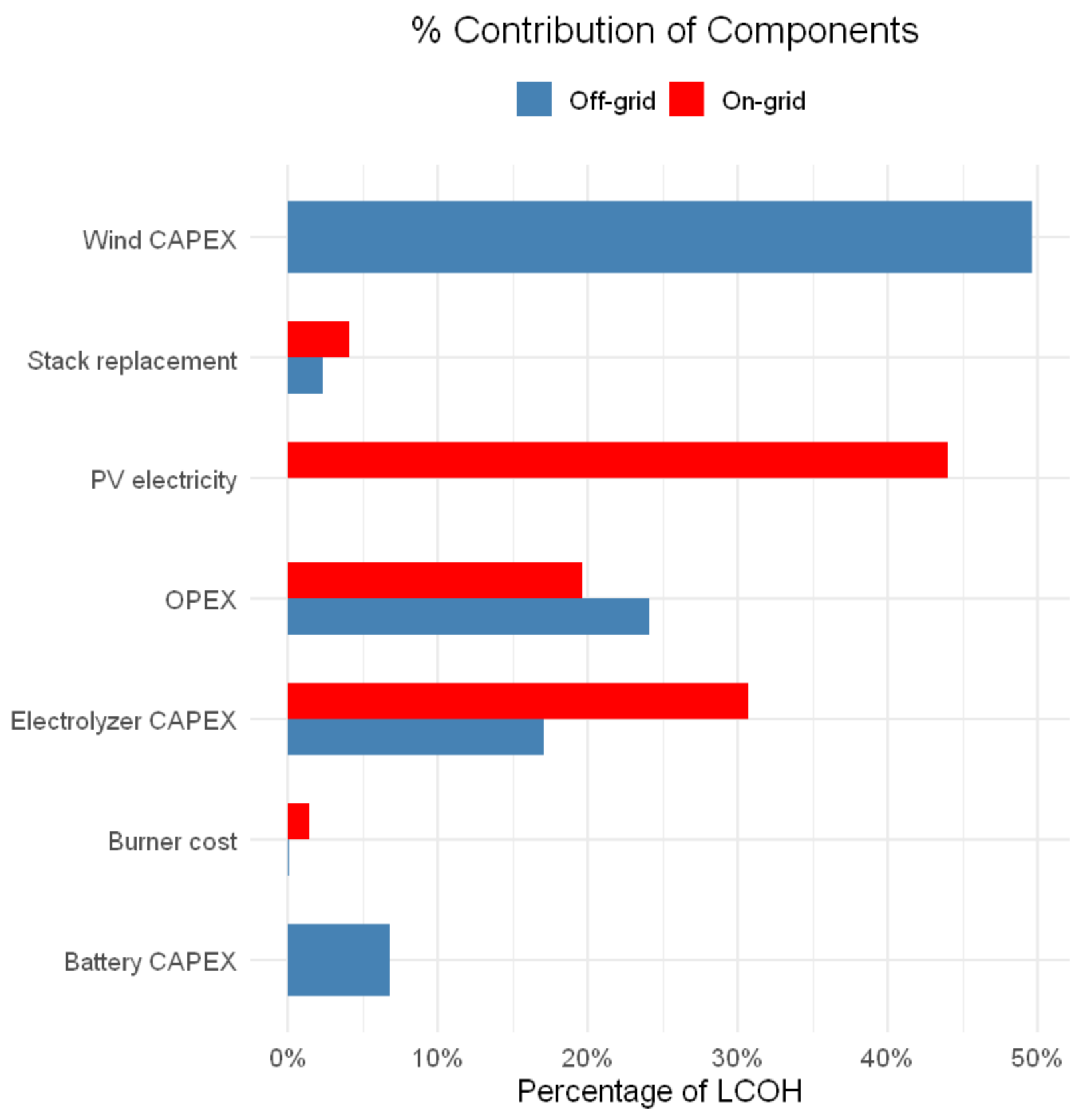

**Fig:6 Scenario Analysis**

**6.5 Carbon Abatement and carbon breakeven analysis**

The figure 7 represents the annual $CO_2$ emissions (expressed in tons of $CO_2$ per year) associated with the thermal demand of five small- and medium-sized manufacturer (SMM) sectors under different levels of natural gas replacement by green hydrogen.The $CO_2$-abatement results, together with the carbon-breakeven assessment, clearly demonstrate the strong decarbonisation potential of green hydrogen as a replacement for natural gas in small- and medium-sized manufacturing sectors. In the Figure , the row labelled "100% $CO_2$ Emission" represents the baseline case, where the entire industrial thermal load is met solely by natural gas combustion (0% hydrogen), while the rows labelled "25% $H_2$ Replacement," "50% $H_2$ Replacement," and "75% $H_2$ Replacement" indicate the remaining $CO_2$ emissions after 25%, 50%, and 75% of the natural-gas energy input (on an LHV basis) has been displaced by green hydrogen. Because hydrogen combustion produces zero process-related $CO_2$, emissions decrease linearly with increasing substitution, reaching zero at 100% hydrogen replacement. For example, in the food sector, baseline emissions of 9,476 $tCO_2$/year fall to 7,107 t, 4,738 t, and 2,369 t at 25%, 50%, and 75% substitution, respectively, corresponding to proportional $CO_2$ avoidance at each level. Across all five sectors, complete substitution eliminates approximately 35,180 $tCO_2$ annually, highlighting the substantial direct emission-reduction potential. Complementing these abatement gains, the carbon-breakeven analysis shows that hydrogen can also be economically favourable. In the on-grid hydrogen scenario, all sectors exhibit negative breakeven carbon prices (e.g., –24 to –23 \$/$tCO_2$ for food; –23.7 to –23.0 \$/$tCO_2$ for glass), indicating that hydrogen reaches cost-parity with natural gas even without any positive carbon price, primarily due to the advantage of wind-rich grid electricity. In the off-grid wind-powered case, breakeven values fall between ~207 and 540 \$/$tCO_2$, reflecting the higher cost of isolated renewable systems but showing clear pathways for improvement as electrolyser and wind-energy costs continue to decline. Taken together, these findings present a highly encouraging outlook: green hydrogen not only provides quantifiable, linear, and complete $CO_2$ elimination when fully substituted for natural gas, but it is also becoming increasingly aligned with the economics of future carbon-constrained industrial energy systems positioning hydrogen as a realistic, scalable, and hopeful route for decarbonisation in SMM sectors.

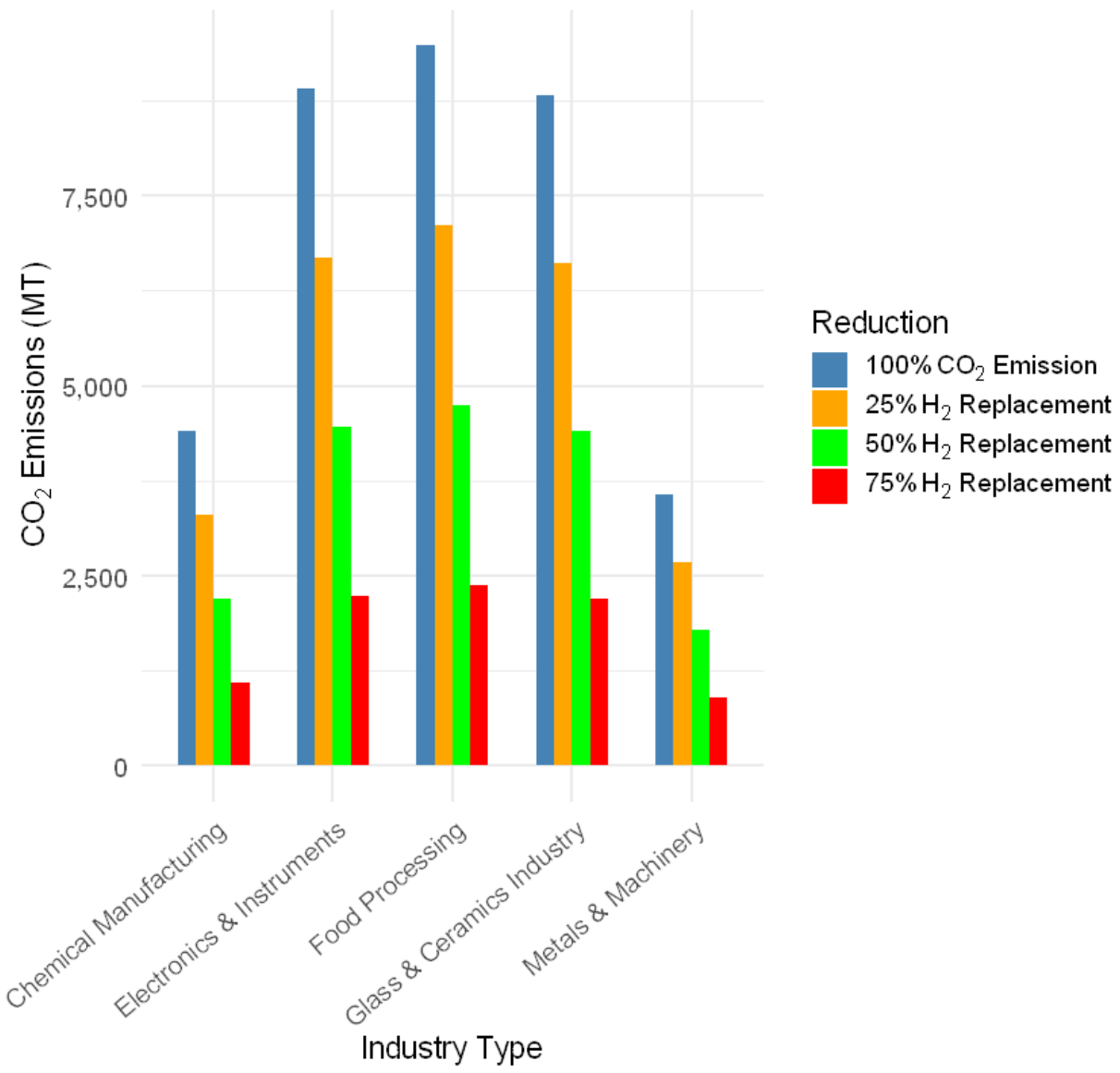

**Fig:7 $CO_2$ Reduction by NG replacement by $H_2$**

## 7.Conclusion

This work developed a unified mixed-integer linear programming framework to evaluate green hydrogen as a substitute for natural gas in high-temperature process heat across different kind of manufacturing industries in USA. The model was applied to two off-grid and on-grid Wind Energy configuration where the electrolyser is supplied energy to produce the hydrogen respecting annual hydrogen production sufficient to replace the targeted natural gas thermal demand in each sector.

Across all industrial cases, the results indicate that off-grid wind-based hydrogen in high-quality wind regions can achieve levelized costs of hydrogen (LCOH) in the range of 8$/kg, depending on sectoral load profile and sector-specific annual operating hours.In the on-grid, LMP-driven configurations show a different trade-off. By exploiting low -price hours the stochastic MILP identifies operational strategies that substantially reduce expected electricity spending, while still satisfying each sector's annual hydrogen requirement. However, even under optimised price-responsive operation, grid-connected hydrogen in our case study remains at LCOH levels of approximately [0.5 $/kg].This is showing that grid-connected projects are highly sensitive to marginal grid emissions and policy design, and that tax incentives and clean-power matching rules will strongly influence their long-term competitiveness.

From a sectoral perspective, the model suggests that industries with high, relatively flat annual thermal loads are particularly promising early adopters. Their high utilisation allows the electrolyser CAPEX to be spread over many full-load hours, driving down LCOH and narrowing the gap to natural gas on a thermal-equivalent basis.

Overall, the results support an optimistic but realistic conclusion: in wind-rich regions such as the U.S. Midwest, wind-to-hydrogen systems serving industrial heat can approach cost ranges that are broadly consistent with forward-looking estimates of green hydrogen, particularly when combined with expected cost declines and supportive policies. Future work should extend this framework to include explicit hydrogen storage options, multi-node power system interactions, and detailed emissions accounting (hourly or sub-hourly matching of clean power), as well as exploring hybrid portfolios that combine off-grid and grid-connected hydrogen supplies across multiple industrial clusters. Together, these extensions will further clarify how green hydrogen can evolve from niche projects into a scalable decarbonization solution for U.S. manufacturing.

**Abbreviations**

| | |
|---|---|
| CAGR | Compound annual growth rate |
| PEM | Proton-exchange-membrane |
| LMP | locational marginal prices |
| LCOH | Levelized Cost of Hydrogen |
| CAPEX | Capital expenditure |
| OPEX | Operational expenditure |
| NG | Natural Gas |
| MILP | Mixed-integer linear programming |
| NPC | Net present cost |
| SMM | Small and Medium-sized Manufacturer |
| BESS | Battery Energy Storage System |
| NREL | National Renewable Energy Laboratory |

**Symbols**

| | |
|---|---|
| $D_j^{\mathrm{H_2}}$ | Demand of hydrogen(kg/yr) |
| $H_t^{\mathrm{prod}}$ | Mass of hydrogen production (kg/hr) |
| $P_{\mathrm{max}}$ | Maximum Power of Electrolyser for on-grid supply(MW) |
| $P^{\mathrm{rated}}$ | Maximum Power of Electrolyser for off-grid supply(MW) |
| $P_t / P_{s,t}$ | Rated Power of Electrolyser in Off gird-Ongrid supply(MW) |
| $E^{\mathrm{batt}}$ | BESS energy capacity.(kWh) |
| $P_{\mathrm{rated}}^{\mathrm{WT}}$ | Installed wind capacity (MW) |
| $P_{\mathrm{rated}}^{\mathrm{batt}}$ | BESS power capacity (MW) |
| HHV | Higher heating value( kWh/kg) |
| $v_{\mathrm{ci}}$ | Cut-in speed(m/s) |
| $v_{\mathrm{r}}$ | Rated Speed(m/s) |
| $v_{\mathrm{co}}$ | Cut out Speed(m/s) |
| $A_{wt}$ | Area of Wind Turbine($m^2$) |
| ch | charging |
| dch | discharging |
| η | Efficiency |
| x | Binary Variable |